\documentclass[10pt,journal,twocolumn]{IEEEtran}
\usepackage[cmex10]{amsmath}
\usepackage{algorithm}
\usepackage{algorithmic}
\usepackage{graphicx} 
\usepackage[varg]{txfonts}
\usepackage{cite}

\usepackage{colortbl}

\ifCLASSINFOpdf
\else
\fi
\begin{document}
\title{Energy Efficiency and Spectral Efficiency Tradeoff in Device-to-Device (D2D) Communications}

\author{Zhenyu~Zhou,~\IEEEmembership{Member,~IEEE,}
        Mianxiong~Dong,~\IEEEmembership{Member,~IEEE,}
        Kaoru~Ota,~\IEEEmembership{Member,~IEEE,}
        Jun~Wu,~\IEEEmembership{Member,~IEEE,}
        and~Takuro~Sato,~\IEEEmembership{Fellow,~IEEE}% <-this % stops a space
 \thanks{Manuscript received May 7, 2014; revised July 4, 2014.}
 \thanks{This work was partially supported by Fundamental Research Funds for the Central Universities under Grant Number 14MS08, China Mobile Communication Co. Ltd. Research Institute (CMRI), and China Electric Power Research Institute (CEPRI) of State Grid Corporation of China (SGCC).}
 \thanks{Zhenyu Zhou is with the State Key Laboratory of Alternate Electrical Power System with Renewable Energy Sources, School of Electrical and Electronic Engineering, North China Electric Power University, Beijing, China, 102206.}
\thanks{Mianxiong Dong is with the National Institute of Information and Communications Technology, Kyoto, Japan. }
\thanks{Kaoru Ota is with the Department of Information and Electric Engineering, Muroran Institute of Technology, Muroran, Hokkaido, Japan (e-mail: ota@csse.muroran-it.ac.jp).}
\thanks{Jun Wu is with the School of Information Security Engineering, Shanghai Jiao Tong University, Shanghai, China.}
\thanks{Takuro Sato is with the Graduate School of Fundamental Science and Engineering, Waseda University, Tokyo, Japan.}}


\markboth{Long Version of IEEE WIRELESS COMMUNICATIONS LETTERS,~Vol.~X, No.~X, January~2014}%
{Shell \MakeLowercase{\textit{et al.}}: Bare Demo of IEEEtran.cls for Journals}



\maketitle

\begin{abstract}
%\boldmath
In this letter, we investigate the tradeoff between energy efficiency (EE) and spectral efficiency (SE) in device-to-device (D2D) communications underlaying cellular networks with uplink channel reuse. The resource allocation problem is modeled as a noncooperative game, in which each user equipment (UE) is self-interested and wants to maximize its own EE. Given the SE requirement and maximum transmission power constraints, a distributed energy-efficient resource allocation algorithm is proposed by exploiting the properties of the nonlinear fractional programming. The relationships between the EE and SE tradeoff of the proposed algorithm and system parameters are analyzed and verified through computer simulations.
\end{abstract}


% \fi
\IEEEpeerreviewmaketitle

\section{Introduction}
\IEEEPARstart{D}{evice-to-device (D2D)} communications underlaying cellular networks bring numerous benefits including the proximity gain, the reuse gain, and the hop gain \cite{D2D_design}. However, the introduction of D2D communications into cellular networks poses many new challenges in the resource allocation design due to the co-channel interference caused by spectrum reuse and limited battery life of user equipments (UEs). 

A large number of works have been done in how to optimize the spectral efficiency (SE) through resource allocation in an interference-limited environment (see \cite{Feiran_WCNC2013, Doppler_TWC, Song_JSAC} and references therein). However, most of the previous studies ignore the energy consumption of UEs. In practical implementation, UEs are typically handheld devices with limited battery life and can quickly run out of battery if the energy consumption is ignored in the system design. 

A limited amount of works have considered the energy efficiency (EE) optimization problem (see \cite{Feiran_2012, Zhou_GC2014v2, Wu_TVT2014}, and references therein). Unfortunately, optimum EE and SE are not always achievable simultaneously and may sometimes even conflict with each other \cite{EE_SE_tradeoff}. Therefore, it is an urgent task to study the EE and SE tradeoff in D2D communications underlaying cellular networks, which has not been well investigated and analyzed. 

In this letter, firstly, we model the resource allocation problem as a noncooperative game, and propose a novel distributed energy-efficient resource allocation algorithm to maximize each UE's EE subject to the SE requirement and transmission power constraints. Then, we study the EE and SE tradeoff of the proposed algorithm, and analyze and verify the relationships between the tradeoff and system parameters (such as transmission power, channel gain, etc.) through computer simulations.

\section{System Model}
\label{System Model}

In this paper, we consider the uplink scenario of a single cellular network. Each cellular UE is allocated with an orthogonal link, and D2D pairs reuse the same channels allocated to cellular UEs in order to improve the SE. The set of UEs is denoted as $\mathcal{S}=\{ \mathcal{N}, \mathcal{K} \}$, where $\mathcal{N}$ and $\mathcal{K}$ denote the sets of D2D UEs and cellular UEs respectively. The total number of D2D links and cellular links are denoted as $N$ and $K$ respectively. 

The distributed resource allocation problem is modeled as a noncooperative game. The strategy sets of the $i$-th D2D transmitter and other D2D transmitters in $\mathcal{N} \backslash \{i\}$ are denoted as $\mathbf{p}_i^d$ and $\mathbf{p}_{-i}^d$ respectively. The strategy sets of the $k$-th cellular UE and other cellular UEs in $\mathcal{K} \backslash \{k\}$ are denoted as $\mathbf{p}_k^c$ and $\mathbf{p}_{-k}^c$ respectively. For the $i$-th D2D pair, its EE $U_{i, EE}^d$ (bits/Hz/J) depends not only on $\mathbf{p}_i^d$, but also on the strategies taken by other UEs in $\mathcal{S}\backslash \{i\}$, i.e.,  $\mathbf{p}_{-i}^d, \mathbf{p}_k^c, \mathbf{p}_{-k}^c$, which is defined as
\begin{align}
\label{eq:UE_EED}
&U_{i, EE}^d (\mathbf{p}_i^d, \mathbf{p}_{-i}^d, \mathbf{p}_k^c, \mathbf{p}_{-k}^c)=\frac{U_{i, SE}^d}{p_{i, total}^d}\notag\\
&=\frac{\sum_{k=1}^K \log_2 \left( 1+\frac{p_i^k g_{i}^k}{p_c^k g^k_{c, i}+\sum_{j=1, j\neq i}^{N}p_{j}^k g_{j, i}^k+N_0} \right) }{\sum_{k=1}^K \frac{1}{\eta } p_i^k+2p_{cir}},
\end{align}
where $U_{i,SE}^d$ is the SE (bits/s/Hz), and $p_{i, total}^d$ is the total power consumption (W). $p_i^k$, $p_c^k$, and $p_{j}^k$ are the transmission power of the $i$-th D2D transmitter, the $k$-th cellular UE, and the $j$-th D2D transmitter in the $k$-th channel respectively. $g_{i}^k$ is the channel gain of the $i$-th D2D pair, $g^k_{c, i}$ is the interference channel gain between the $k$-th cellular UE and the $i$-th D2D receiver, and $g_{j, i}^k$ is the interference channel gain between the $j$-th D2D transmitter and the $i$-th D2D receiver. $p_c^k g^k_{c, i}$ and $\sum_{j=1, j\neq i}^{N} p_{j}^k g_{j, i}^k$ denote the interference from the cellular UE and the other D2D pairs that reuse the $k$-th channel respectively. $N_0$ is the noise power. $p_{i, total}^d$ is composed of the transmission power over all of the $K$ channels, i.e., $\sum_{k=1}^K \frac{1}{\eta } p_i^k$, and the circuit power of both the D2D transmitter and receiver, i.e., $2p_{cir}$. The circuit power of any UE is assumed as the same and denoted as $p_{cir}$. $\eta$ is the power amplifier (PA) efficiency, i.e., $0 < \eta < 1$.

Similarly, the EE of the $k$-th cellular UE $U_{k, EE}^c$ is defined as
\begin{align}
&U_{k, EE}^c (\mathbf{p}_i^d, \mathbf{p}_{-i}^d, \mathbf{p}_k^c, \mathbf{p}_{-k}^c)\notag\\
&=\frac{U_{k, SE}^c}{p_{k, total}^c}=\frac{\log_2 \left( 1+\frac{p_c^k g_c^k}{\sum_{i=1}^{N}p_{i}^k g_{i, c}^k+N_0} \right)}{\frac{1}{\eta} p_c^k+p_{cir}},
\end{align}
where $g_c^k$ is the channel gain between the $k$-th cellular UE and the base station (BS), $g^k_{i, c}$ is the interference channel gain between the $i$-th D2D transmitter and the BS in the $k$-th channel. $\sum_{i=1}^{N}p_{i}^k g_{i, c}^k$ denotes the interference from all of the D2D pairs to the BS in the $k$-th channel. $p_{k, total}^c$ is composed of the transmission power $\frac{1}{\eta} p_c^k$ and the circuit power only at the transmitter side, i.e., $p_{cir}$.  

The EE maximization problem for the $i$-th D2D pair is formulated as
\begin{align}
\label{eq:Dproblem}
&\max. \hspace{5mm} U_{i, EE}^d (\mathbf{p}_i^d, \mathbf{p}_{-i}^d, \mathbf{p}_k^c, \mathbf{p}_{-k}^c) \\
&\mbox{s.t.} \hspace{5mm} C1: U_{i, SE}^d  \geq R_{i, min}^d, \\
 &\hspace{9mm} C2: 0 \leq \sum_{k=1}^K p_i^k \leq p_{i, max}^d.
\end{align}

The corresponding EE maximization problem for the $k$-th cellular UE is formulated as
\begin{align}
\label{eq:Cproblem}
&\max. \hspace{5mm} U_{k, EE}^c (\mathbf{p}_i^d, \mathbf{p}_{-i}^d, \mathbf{p}_k^c, \mathbf{p}_{-k}^c) \\
&\mbox{s.t.} \hspace{5mm} C3: U_{k, SE}^c  \geq R_{k, min}^c,\\
 &\hspace{9mm} C4: 0 \leq p_c^k \leq p_{k, max}^c.
\end{align}
The constraints C1 and C3 specify the minimum SE requirements. C2 and C4 are the non-negative constraints on the power allocation variables.

\section{Distributed Energy-Efficient Resource Allocation}
\label{distributed}

\subsection{The Objective Function Transformation}
\label{transformation}
The objective functions defined in (\ref{eq:Dproblem}) and (\ref{eq:Cproblem}) are non-convex, but can be transformed into concave functions by using the nonlinear fractional programming developed in \cite{Dinkelbach}. 
We define the maximum EE of the $i$-th D2D pair as $q^{d*}_i$, which is given by
\begin{equation}
q^{d*}_i=\max. \hspace{1mm} U_{i, EE}^d (\mathbf{p}_i^d, \mathbf{p}_{-i}^d, \mathbf{p}_k^c, \mathbf{p}_{-k}^c)=\frac{U_{i, SE}^d(\mathbf{p}_i^{d*})}{p_{i, total}^d(\mathbf{p}_i^{d*})},
\end{equation} 
where $\mathbf{p}_i^{d*}$ is the best response of the $i$-th D2D transmitter given the other UEs' strategies $\mathbf{p}_{-i}^d$, $\mathbf{p}_k^c$, $\mathbf{p}_{-k}^c$. The following theorem can be proved:

\textbf{\emph{Theorem 1:}} The maximum EE $q_i^{d*}$ is achieved if and only if $\max. \:\: U_{i, SE}^d (\mathbf{p}_i^d)-q_i^{d*}p_{i, total}^d(\mathbf{p}_i^d)=U_{i,SE}^d (\mathbf{p}_i^{d*})-q_i^{d*}p_{i, total}^d(\mathbf{p}_i^{d*})=0$.
  \begin{IEEEproof}
The proof of Theorem 1 is given in Appendix \ref{theorem1}.
\end{IEEEproof}

Theorem 1 shows that the transformed problem with an objective function in subtractive form is equivalent to the non-convex problem in fractional form, i.e., they lead to the same optimum solution $\mathbf{p}_i^{d*}$. Similarly, let $q^{c*}_k$ and $\mathbf{p}_k^{c*}$ denote the maximum EE and best response of the $k$-th cellular UE, we have 

\textbf{\emph{Theorem 2:}} The maximum EE $q_k^{c*}$ is achieved if and only if $\max. \:\: U_{k, SE}^c (\mathbf{p}_k^c)-q_k^{c*}p_{k, total}^c(\mathbf{p}_k^c)=U_{k, SE}^c (\mathbf{p}_k^{c*})-q_k^{c*}p_{k, total}^c(\mathbf{p}_k^{c*})=0$.

\subsection{The Iterative Optimization Algorithm}
\label{algorithm} 
The proposed algorithm is summarized in Algorithm \ref{offline algorithm}. $n$ is the iteration index, $L_{max}$ is the maximum number of iterations, and $\Delta$ is the maximum tolerance. $L_{max}$ is set to 10 to ensure that the algorithm converges sufficiently although simulation results in Section \ref{Simulation Results} show that the algorithm is able to converge in only 5 iterations. This setting will not increase the computation complexity significantly because the loop will terminate once the algorithm converges sufficiently close to the optimum EE, i.e., when the condition $U_{i, SE}^d(\mathbf{\hat{p}}_i^d)-q_i^d p_{i,total}^d (\mathbf{\hat{p}}_i^d) \leq \Delta$ is satisfied.

At each iteration, for any given $q_i^{d}$ or $q_k^{c}$, the corresponding resource allocation strategies are obtained by solving the following equivalent transformed optimization problems respectively:
\begin{align}
 \label{eq:transformed problemD}
  &\max . \:\: U_{i, SE}^d (\mathbf{p}_i^d)-q_i^d p_{i, total}^d(\mathbf{p}_i^d) \nonumber\\
 &\mbox{s.t.} \:\:\: C1, C2.
 \end{align}
 \begin{align}
 \label{eq:transformed problemC}
  &\max . \:\: U_{k, SE}^c (\mathbf{p}_k^c)-q_k^c p_{k, total}^c (\mathbf{p}_k^c) \nonumber\\
 &\mbox{s.t.} \:\:\: C3, C4.
 \end{align}
 
  Taking the $i$-th D2D pair as an example, the Lagrangian associated with the problem (\ref{eq:transformed problemD}) is given by
\begin{align}
&\mathcal{L}_{EE}(\mathbf{p}_i^d, \alpha_i, \beta_i) =U_{i, SE}^d  (\mathbf{p}_i^d)-q_i^d p_{i, total}^d (\mathbf{p}_i^d) \notag\\
&+\alpha_i \left( U_{i, SE}^d(\mathbf{p}_i^d)-R_{i, min}^d \right)-\beta_i \left( \sum_{k=1}^K p_i^k-p_{i, max}^d\right),
 \end{align}
 where $\alpha_i$, $\beta_i$ are the Lagrange multipliers associated with the constraints C1 and C2 respectively. Since the problem (10) is in a standard concave form with differentiable objective and constraint functions, the Karush-Kuhn-Tucker (KKT) conditions are used to find the optimum solutions and the duality gap is zero (see page 244 in \cite{convex_optimization}). 
Another way to prove that the strong duality holds is to prove that the Slater's condition is satisfied. Define $f_0 (\mathbf{p}_i^d)=-U_{i, SE}^d (\mathbf{p}_i^d)+q_i^d p_{i, total}^d(\mathbf{p}_i^d)$, $f_1 (\mathbf{p}_i^d)=R_{i, min}^d-U_{i, SE}^d(\mathbf{p}_i^d)$, $f_2 (\mathbf{p}_i^d)= - \sum_{k=1}^K p_i^k$, $f_3 (\mathbf{p}_i^d)=\sum_{k=1}^K p_i^k-p_{i, max}^d$, then the EE maximization problem can be written as
\begin{align}
&\min. \hspace{5mm} f_0 (\mathbf{p}_i^d) \\
&\mbox{s.t.} \hspace{5mm} f_1 (\mathbf{p}_i^d) \leq 0 \\
&\hspace{9mm} f_2 (\mathbf{p}_i^d) \leq 0 \\
&\hspace{9mm}  f_3(\mathbf{p}_i^d) \leq 0 
\end{align}

Let us define \textbf{relint} $\mathcal{D}$ as the relative interior of the feasible domain, and $\mathcal{D}=\cap_{m=1}^{3} \mbox{dom} (f_m)$. We note that $f_0$ and $f_1$ are convex functions, and $f_2$ and $f_3$ are affine functions. If \textbf{relint} $\mathcal{D}$ is not empty, there always exists an $\mathbf{p}_i^d \in $  \textbf{relint} $\mathcal{D}$ such that $f_1 (\mathbf{p}_i^d) < 0$, which satisfies the Slater's condition and ensures that the strong duality holds. On the other hand, if \textbf{relint} $\mathcal{D}$ is empty, the optimization problem is either infeasible or has only one solution, which is not the interest of this paper.

Alternatively, we can replace $R_{i, min}^d$ by $R_{i, min}^d+\lim_{\xi \to 0^{+}} \xi $ ($\xi >0$) in the constraint C1 so that $U_{i, SE}^d(\mathbf{p}_i^d) \geq R_{i, min}^d+\lim_{\xi \to 0^{+}} \xi$. This always ensures that
\begin{align}
 f_1=R_{i, min}^d-U_{i, SE}^d(\mathbf{p}_i^d) \leq R_{i, min}^d-R_{i, min}^d-\lim_{\xi \to 0^{+}} \xi=-\lim_{\xi \to 0^{+}} \xi <0. 
\end{align}
This modification of C2 will not affect the stability of the algorithm since the proposed iterative optimization algorithm converges to the optimum EE, which is proved in Theorem 4. 

The equivalent dual problem can be decomposed into two subproblems, which is given by 
 \begin{equation}
\label{eq:dual problem}
 \displaystyle \min_{\displaystyle (\alpha_i \geq 0, \beta_i \geq 0)}\!\!\!\!. \hspace{5mm} \max_{\displaystyle (\mathbf{p}_i^d)}. \:\:\: \mathcal{L}_{EE}(\mathbf{p}_i^d, \alpha_i, \beta_i) 
\end{equation}
Taking the first-order derivatives of (12) with regard to $p_i^k$, we have
\begin{align}
\frac{\partial \mathcal{L}_{EE}(\mathbf{p}_i^d, \alpha_i, \beta_i)}{\partial p_i^k} \Big |_{p_i^k=\hat{p}_i^k}=0, k=1, \cdots, K
\end{align}
 For any given $q_i^d$, the optimum solution is given by
\begin{equation}
\label{eq:waterfilling}
\hat{p}_i^{k}=\left[ \frac{\eta(1+\alpha_i) \log_2e }{q_i^d+\eta\beta_i }-\frac{\hat{p}_c^k g^k_{c, i}+\sum_{j=1, j\neq i}^{N}\hat{p}_{j}^k g_{j, i}^k+N_0}{g_{i}^k}\right]^{+},
\end{equation}
where $[x]^+=\max\{0,x\}$. Equation (\ref{eq:waterfilling}) indicates a water-filling algorithm for transmission power allocation, and the interference from the other UEs decreases the water level. 

For solving the minimization problem, the Lagrange multipliers can be updated by using the gradient method \cite{improved_step_size, subgradient}. The gradient of $\alpha_i$ and $\beta_i$ are given by
\begin{align}
&\frac{\partial \mathcal{L}_{EE}(\mathbf{p}_i^d, \alpha_i, \beta_i)}{\partial \alpha_i}=U_{i, SE}^d(\mathbf{p}_i^d)-R_{i, min}^d, \notag\\
&\frac{\partial \mathcal{L}_{EE}(\mathbf{p}_i^d, \alpha_i, \beta_i)}{\partial \beta_i}=-\left( \sum_{k=1}^K p_i^k-p_{i, max}^d \right).
\end{align}
Then, $\alpha_i$, $\beta_i$ are updated by using the gradient method as
\begin{align}
\alpha_i (\tau +1)&=\left[ \alpha_i(\tau )-\mu_{i, \alpha} (\tau ) \left(U_{i, SE}^d(\mathbf{\hat{p}}_i^d)-R_{i, min}^d \right)    \right]^{+},\\
\beta_i (\tau +1)&=\left[ \beta_i(\tau )+\mu_{i, \beta} (\tau ) \left( \sum_{k=1}^K \hat{p}_i^k  - p_{i, max}^d \right)    \right]^{+},
\end{align}
where $\tau \geq 0$ is the iteration index, $\mu_{i, \alpha} (\tau ) , \mu_{i, \beta} (\tau ) $ are the positive step sizes which are taken in the direction of the negative gradient for the dual variables at iteration $\tau$. The step sizes should be chosen to strike a balance between optimality and convergence speed. Since the Lagrange multiplier updating techniques are beyond the scope of this paper, interested readers may refer to \cite{improved_step_size, subgradient} and references therein for details.

 %as
%\begin{align}
%\alpha_i (\tau +1)&=\left[ \alpha_i(\tau )-\mu_{i, \alpha} (\tau ) \left( r_i^d(\tau )-R_{i, min}^d \right)    \right]^{+},\\
%\beta_i (\tau +1)&=\left[ \beta_i(\tau )-\mu_{i, \beta} (\tau ) \left( \sum_{k=1}^K p_i^k (\tau) - p_{i, max}^d \right)    \right]^{+},
%\end{align}
%where $\tau$ is the iteration index, $\mu_{i, \alpha}, \mu_{i, \beta}$ are the positive step sizes. The solution of problem (\ref{eq:dual problem}) converges to the optimum solution in (\ref{eq:transformed problemD}) if the step sizes are chosen to satisfy the diminishing step size rules \cite{subgradient}. 
Similarly, for any given $q_k^c$, the optimum solution of $k$-th cellular UE is given by
\begin{equation}
\label{eq:waterfilling_CE}
\hat{p}_c^k=\left[ \frac{\eta (1+\delta_k) \log_2e }{q_k^c+\eta \theta_k  }-\frac{\sum_{i=1}^N \hat{p}_{i}^k g_{i, c}^k+N_0}{g_{c}^k} \right]^+,
\end{equation}
where $\delta_k, \theta_k$ are the Lagrange multipliers associated with the constraints C3 and C4 respectively.

\subsection{Complexity Analysis}

The proposed iterative optimization algorithm is based on the nonlinear fractional programming developed in \cite{Dinkelbach}. The iterative algorithm solves the convex problem of (\ref{eq:transformed problemD}) (or (\ref{eq:transformed problemC}) at each iteration. The iterative algorithm produces an increasing sequence of $q_i^d$ (or $q_k^c$) values which are proved to converge to the optimum EE $q_i^{d*}$ at a superlinear convergence rate \cite{Dinkelbach_superlinear}. Taking the $i$-th D2D pair as an example, in each iteration, (\ref{eq:transformed problemD}) is solved by using the Lagrange dual decomposition. The algorithmic complexity of this method is dominated by the calculations given by (\ref{eq:waterfilling}), which leads to a total complexity $\mathcal{O} (I_{i, dual}^d I_{i, loop}^{d}  K)$ when $K$ is large, where $I_{i, dual}^d$ is the required number of iterations required for reaching convergence, i.e., $I_{i, dual} \leq L_{max}$, and $I_{i, loop}^{d}$ is the required number of iterations for solving the dual problem. 

In particular, the dual problem (\ref{eq:dual problem}) is decomposed into two subproblems: the inner maximization problem solves the the power allocation problem to find the best strategy and the outer minimization problem solves the master dual problem to find the corresponding Lagrange multipliers. In the inner maximization problem, a total of $I_{i, dual}^d  I_{i, loop}^{d}K (N+3)$ real additions, $I_{i, dual}^d I_{i, loop}^{d}K(N+5)$ real multiplications, and $I_{i, dual}^d I_{i, loop}^{d}K$ real comparisons are required. In the outer minimization problems, a total of $I_{i, dual}^d I_{i, loop}^{d}(K+3)$ real additions, $2 I_{i, dual}^d I_{i, loop}^{d}$ real multiplications, and $2 I_{i, dual}^d I_{i, loop}^{d}$ real comparisons are quired. In conclusion, a total of $I_{i, dual}^d I_{i, loop}^{d}(KN+4K+3)$ real additions, $I_{i, dual}^d I_{i, loop}^{d}(KN+5K+2)$ real multiplications, and $I_{i, dual}^d I_{i, loop}^{d}(K+2)$ real comparisons are quired for the $i$-th D2D pair.

\begin{algorithm}[t]
\caption{Iterative Resource Allocation Algorithm}
\label{offline algorithm}
\begin{algorithmic}[1]
\STATE $q_i^d \leftarrow 0$, $q_k^c \leftarrow 0$, $L_{max} \leftarrow 10$, $n \leftarrow 1$, $\Delta \leftarrow 10^{-3}$ 
\FOR{$n=1$ to $L_{max}$}
\IF {D2D link}
\STATE solve (\ref{eq:transformed problemD}) for a given $q_i^d$ and obtain $\mathbf{\hat{p}}_i^d$
\IF{$U_{i, SE}^d(\mathbf{\hat{p}}_i^d)-q_i^d p_{i,total}^d (\mathbf{\hat{p}}_i^d) \leq \Delta$,}
 \STATE $\mathbf{p}_i^{d*}=\mathbf{\hat{p}}_i^d$, and $\displaystyle q_i^{d*}=\frac{U_{i,SE}^d(\mathbf{p}_i^{d*})}{p_{i, total}^d(\mathbf{p}_i^{d*})}$ 
\STATE \textbf{break}
\ELSE
\STATE $\displaystyle q_i^d=\frac{U_{i,SE}^d(\mathbf{\hat{p}}_i^d)}{p_{i, total}^d(\mathbf{\hat{p}}_i^d)}$, and $n=n+1$
\ENDIF
\ELSE
\STATE solve (\ref{eq:transformed problemC}) for a given $q_k^c$ and obtain $\mathbf{\hat{p}}_k^c$
\IF{$U_{k,SE}^c(\mathbf{\hat{p}}_k^c)-q_k^c p_{k,total}^c (\mathbf{\hat{p}}_k^c) \leq \Delta$,}
 \STATE $\mathbf{p}_k^{c*}=\mathbf{\hat{p}}_c$, and $\displaystyle q_k^{c*}=\frac{U_{k,SE}^c (\mathbf{p}_k^{c*})}{p_{k, total}^c(\mathbf{p}_k^{c*})}$ 
\STATE \textbf{break}
\ELSE
\STATE $\displaystyle q_k^c=\frac{U_{k,SE}^c(\mathbf{\hat{p}}_k^c)}{p_{k, total}^c(\mathbf{\hat{p}}_k^c)}$, and $n=n+1$
\ENDIF
\ENDIF
\ENDFOR
\end{algorithmic}
\end{algorithm}

\subsection{Distributed Implementation}

In the formulated EE maximization problem, the best response of the $i$-th D2D transmitter $\mathbf{p}_i^d$ depends on the strategies of all other UEs, i.e., $\mathbf{p}_{-i}^d, \mathbf{p}_k^c, \mathbf{p}_{-k}^c$. In order to obtain this knowledge, each UE has to broadcast its transmission strategy to other UEs. However, we observe that the sufficient information of $\mathbf{p}_{-i}^d, \mathbf{p}_k^c, \mathbf{p}_{-k}^c$ are contained in the form of interference, i.e., $p_c^k g^k_{c, i}$ and $\sum_{j=1, j\neq i}^{N} p_{j}^k g_{j, i}^k$. In this way, each D2D pair has only to estimate the interference on all available channels to determine the power optimization rather than knowing the specific strategies of other UEs. For the $k$-th cellular UE, the BS estimates the interference from D2D pairs on the $k$-th channel and then feeds back this information to the cellular UE. If UEs update their strategies sequentially, player strategies will eventually converge to a Nash equilibrium, which is proved to exist in Theorem 3. The D2D peer discovery techniques and the design of strategy updating mechanism are out of the scope of this paper and will be discussed in future works.

\section{Energy Efficiency and Spectral Efficiency Tradeoff}
\label{tradeoff}

For the $i$-th D2D pair, by analyzing the EE and SE relationships, we have the following properties.

\textbf{\emph{Lemma 1:}} 
The SE, $U_{i, SE}^{d}$,  increases monotonically as $p_i^k$ increases, while the EE, $U_{i, EE}^{d}$, increases firstly and then decreases as $p_i^k$ increases. $U_{i, EE}^d$ is quasiconcave.
  \begin{IEEEproof}
The proof of Lemma 1 is given in Appendix \ref{lemma1}.
\end{IEEEproof}

\textbf{\emph{Lemma 2:}} The transformed objective function in subtractive form is a concave function.
  \begin{IEEEproof}
The proof of Lemma 2 is given in Appendix \ref{lemma2}.
\end{IEEEproof}

\textbf{\emph{Lemma 3:}} $  \max_{(\mathbf{p}_i^d)} U_{i, SE}^d (\mathbf{p}_i^d)-q_i^d p_{i, total}^d(\mathbf{p}_i^d)$ is monotonically decreasing as $q_i^d$ increases. 
  \begin{IEEEproof}
The proof of Lemma 3 is given in Appendix \ref{lemma3}.
\end{IEEEproof}

 \textbf{\emph{Lemma 4:}} For any feasible $\mathbf{p}_i^{d}$, $\max_{\big(\mathbf{p}_i^{d}\big)} U_{i, SE}^d \big(\mathbf{p}_i^{d}\big)-q_i^{d} p_{i, total}^d(\mathbf{p}_i^d ) \geq 0$.
   \begin{IEEEproof}
The proof of Lemma 4 is given in Appendix \ref{lemma4}.
\end{IEEEproof}

  \textbf{\emph{Theorem 3:}} A Nash equilibrium exists and the optimum strategy set $\{ \mathbf{p}_i^{d*}, \mathbf{p}^{c*}_k \mid  i \in \mathcal{N},  k \in \mathcal{K}\}$ obtained by using Algorithm \ref{offline algorithm} is the Nash equilibrium.
  \begin{IEEEproof}
The proof of Theorem 3 is given in Appendix \ref{theorem3}.
\end{IEEEproof}

  \textbf{\emph{Theorem 4:}} The proposed iterative optimization algorithm converges to the optimum EE.
  \begin{IEEEproof}
The proof of Theorem 4 is given in Appendix \ref{theorem4}.
\end{IEEEproof}

 \textbf{\emph{Corollary 1:}} 
EE can be increased by a maximum of $\Delta EE=q_i^{d*}-U_{i, EE}^{d}(\mathbf{p}_i^d)$ by either trading off SE with $\Delta SE= U_{i, SE}^{d}(\mathbf{p}_i^{d})-U_{i, SE}^{d}(\mathbf{p}_i^{d*})$ if and only if
$p_i^k>p_i^{k*}$, $\forall p_i^{k*} \in \mathbf{p}_i^{d*}$, or by simultaneously increasing SE with $\Delta SE= U_{i, SE}^{d}(\mathbf{p}_i^{d*})-U_{i, SE}^{d}(\mathbf{p}_i^{d})$ if and only if $p_i^k<p_i^{k*}$, $\forall p_i^{k*} \in \mathbf{p}_i^{d*}$.
\begin{IEEEproof}
Corollary 1 can be easily proved by Lemma 1 since that $U_{i,EE}^d$ decreases as $p_i^k$ increases when $p_i^k>p_i^{k*}$, and both $U_{i,EE}^d$ and $U_{i,SE}^d$ increases as $p_i^k$ increases when $p_i^k<p_i^{k*}$.
\end{IEEEproof} 

The EE and SE tradeoffs depend on the specific channel realization in each simulation and a large number of simulations are required to obtain the average result. In order to facilitate analysis and get some insights, we consider a special case that all the signal channels  have the same power gain $g$, and all the interference channels have the same power gain $\hat{g}$. 
The network coupling factor is defined as $ I=\hat{g}/g$ \cite{Miao_TWC2011}. Assuming that $N_0$ can be ignored comparing to the interference, $U_{i,SE}^d$ and $U_{i,EE}^d$ are given by
\begin{align}
\label{eq:U_SE_D_I}
U_{i,SE}^d &\approx K \log_2 \big(1+ \frac{p_i^k}{p_c^kI+(N-1)p_i^kI} \big),\\
\label{eq:U_EE_D_I}
U_{i,EE}^d &\approx \frac{\eta U_{i,SE}^d \big( 1-(N-1) I (2^{ \frac{U_{i,SE}^d}{K}}-1) \big)}{K p_c^k I (2^{\frac{U_{i,SE}^d}{K}}-1)+2p_{cir}\eta \big( 1-(N-1) I (2^{\frac{U_{i,SE}^d}{K}}-1) \big) }.
\end{align}

Similarly, $U_{k, SE}^c$ and $U_{i, EE}^d$ are given by
\begin{align}
\label{eq:U_SE_C_K}
U_{k,SE}^c &\approx  \log_2 \big(1+ \frac{p_c^k}{Np_i^kI} \big),\\
\label{eq:U_EE_C_K}
U_{k,EE}^c &\approx \frac{\eta U_{k,SE}^c }{Np_i^kI(2^U_{k,SE}-1)+p_{cir}\eta}.
\end{align}

  \textbf{\emph{Corollary 2:}}
For any given $p_i^k$ and $p_c^k$, both $U_{i,SE}^d$ and $U_{i,EE}^d$ decrease monotonically as $I$ increases. For any finite and positive $I$, $U_{i,EE}^d$ increases firstly and then decreases as $U_{i,SE}^d$ increases. $U_{i,EE}^d \rightarrow  0$ if and only if $U_{i,SE}^d \rightarrow 0$ or $U_{i,SE}^d \rightarrow K\log_2 \big(1+\frac{1}{(N-1)I} \big)$. 
  \begin{IEEEproof}
The proof of Corollary 2 is given in Appendix \ref{corollary2}.
\end{IEEEproof}

  \textbf{\emph{Corollary 3:}}
For any given $p_i^k$ and $p_c^k$, both $U_{k,SE}^c$ and $U_{k,EE}^c$ decrease monotonically as $I$ increases. For any finite and positive $I$, $U_{k,EE}^c$ increases firstly and then decreases as $U_{k,SE}^c$ increases. $U_{k,EE}^c \rightarrow  0$ if and only if $U_{k,SE}^c \rightarrow 0$ or $U_{k,SE}^c \rightarrow \infty$.

Similar conclusions hold for cellular links but are omitted here due to space limitation.

\section{Simulation Results}
\label{Simulation Results}

\begin{figure}[t]
\begin{center}
\scalebox{0.55} 
{\includegraphics{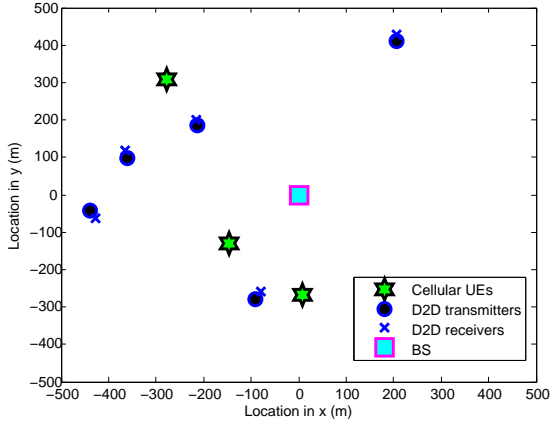}}
\end{center}
\caption{The locations of D2D UEs and cellular UEs generated in one simulation ($N=5$, $K=3$, the cell radius is $500$ m, and maximum D2D distance is $25$ m ).}
\label{scenario}
\end{figure}

In this section, the EE and SE tradeoff is investigated through computer simulations. There are a total of $N=5$ D2D links and $K=3$ cellular links. For each simulation, the locations of cellular UEs and D2D UEs are generated randomly within a cell with a radius of $500$ m. The maximum D2D transmission distance is $25$ m. The values of simulation parameters and channel gains are inspired by \cite{Feiran_WCNC2013, Feiran_2012, Song_JSAC}. Fig. \ref{scenario} shows the locations of D2D UEs and cellular UEs generated in one simulation. The maximum distance between any two D2D UEs that form a D2D pair is $25$ m. The channel gain between the transmitter $i$ and the receiver $j$ is calculated as $d_{i, j}^{-2} |h_{i, j}|^2$ \cite{Feiran_WCNC2013, Feiran_2012, Feiran_ICC2013}, where $d_{i, j}$ is the distance between the transmitter $i$ and the receiver $j$, $h_{i, j}$ is the complex Gaussian channel coefficient that satisfies $h_{i, j} \sim \mathcal{CN} (0, 1)$. 

\begin{figure}[t]
\begin{center}
\scalebox{0.55} 
{\includegraphics{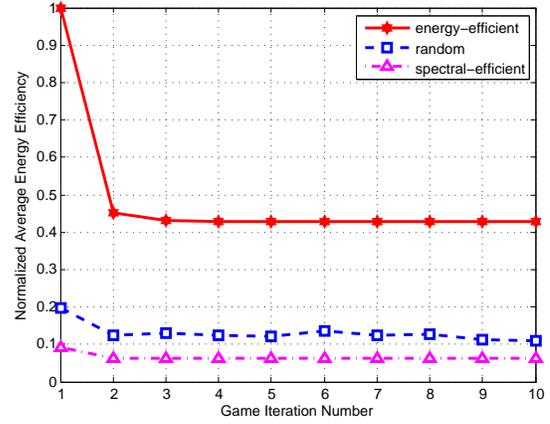}}
\end{center}
\caption{The normalized average energy efficiency of D2D links corresponding to the number of game iterations ($N=5$, $K=3$, $p_{i, max}^d=p_{k, max}^c=200$ mW, $R_{k, min}^c=0.1$ bit/s/Hz, $R_{i, min}^d=1$ bit/s/Hz, $1000$ simulations).}
\label{EE_D2D}
\end{figure}

Fig. \ref{EE_D2D} shows the normalized average EE of D2D links corresponding to the number of game iterations. We compare the proposed EE maximization algorithm (labeled as ``energy-efficient") with the SE maximization algorithm (labeled as ``spectral-efficient" ), and the random power allocation algorithm (labeled as ``random"). In the spectral-efficient algorithm, each UE is self-interested and wants to maximize its own SE rather than EE, and the power consumption is completely ignored in the optimization process. The results are averaged through a total number of $1000$ simulations and  normalized by the maximum value. The normalized average EE of the proposed energy-efficient algorithm converge to $0.429$, while the random algorithm converge to $0.124$ and the spectral-efficient algorithm converge to $0.064$. It is clear that the proposed energy-efficient algorithm significantly outperforms the spectral-efficient algorithm and the random algorithm in terms of EE in an interference-limited environment. The spectral-efficient algorithm has the worst EE performance among the three because power consumption is completely ignored in the optimization process. The random algorithm fluctuates around the equilibrium since that the transmission power strategy is randomly selected.

\begin{figure}[t]
\begin{center}
\scalebox{0.55} 
{\includegraphics{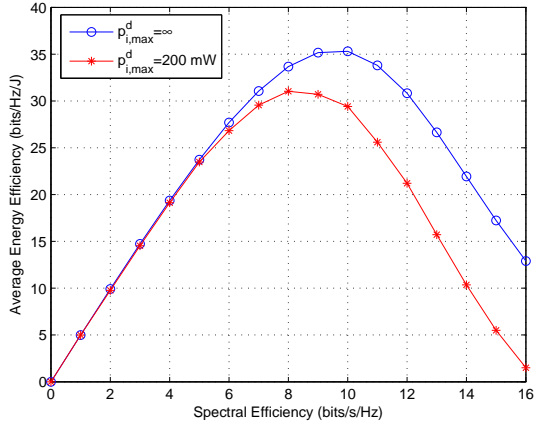}}
\end{center}
\caption{The energy efficiency and spectral efficiency tradeoff for D2D links corresponding to $p_{i,max}^d=\infty$, $200$ mW, ($N=5, K=3$, $N_0=\displaystyle 10^{-7}$ W, $p_{k, max}^c=200$ mW, $\eta=0.35$, $p_{cir}=100$ mW).}
\label{EE_SE_real}
\end{figure}

Fig. \ref{EE_SE_real} shows the EE and SE tradeoffs for D2D links corresponding to $p_{i,max}^d=\infty, 200$ mW respectively. For each D2D link, the SE requirement is increased from $0$ to $16$ bits/s/Hz with a step of $1$,  and the corresponding EE is obtained by Algorithm 1. The average EE of $N$ D2D links is averaged again over a total number of $500$ simulations. For any specified SE requirement ($0 \leq U_{i, EE}^d \leq 16$ bits/s/Hz), there is always a possibility to satisfy the SE requirement if the signal channel gain is large enough compared to the interference channel gain. One simple example is that the $i$-th D2D transmitter and receiver are close to each other but far from the other interference sources. Simulation results show that the maximum achievable EE is limited by $p_{i,max}^d$ (constraint C2), which is particularly obvious in the high SE regime. If the circuit power consumption $p_{cir}$ is taken into consideration, as proved in Lemma 1, the EE, $U_{i, EE}^{d}$, increases firstly and then decreases as $p_i^k$ increases. Since the SE, $U_{i, SE}^{d}$, increases monotonically as $p_i^k$ increases, we can prove that the EE, $U_{i, EE}^{d}$, increases firstly and then decreases as $U_{i, SE}^{d}$ increases, which is in agreement with Fig. \ref{EE_SE_real}. It is clear that the EE gain achieved by decreasing the transmission power below the power for optimum EE is not able to compensate for the EE loss caused by the circuit power and SE loss.

\begin{figure}[t]
\begin{center}
\scalebox{0.55} 
{\includegraphics{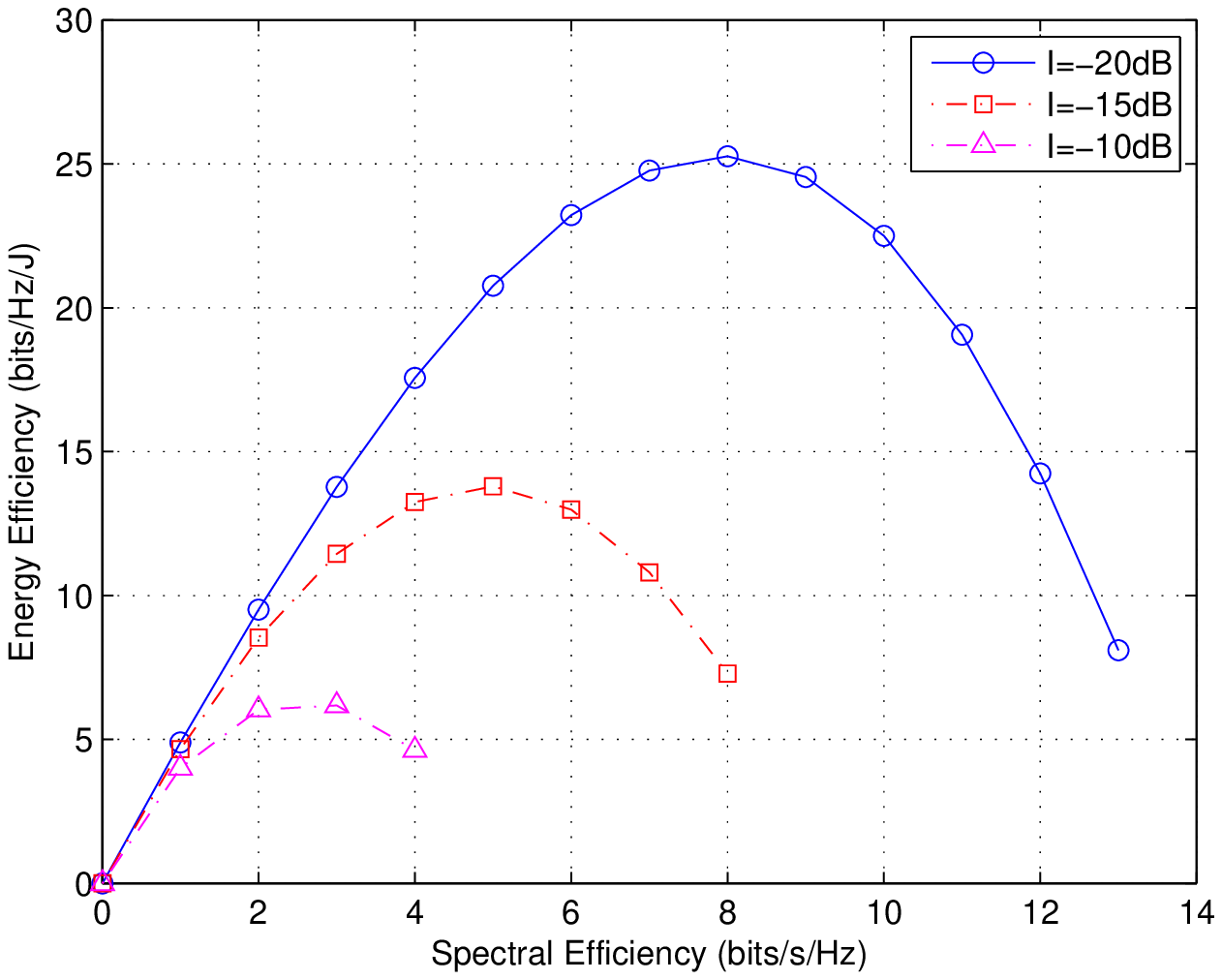}}
\end{center}
\caption{The energy efficiency and spectral efficiency tradeoff for D2D links corresponding to three interference levels $I=-20, -15, -10$ dB, ($g=1, N=5, K=3$, $p_{i, max}^d=p_{k, max}^c=200$ mW, $\eta=0.35$, $p_{cir}=100$ mW).}
\label{EE_SE_tradeoff}
\end{figure}

 Fig. \ref{EE_SE_tradeoff} shows the tradeoff between EE and SE for D2D links in the special case discussed in Section \ref{tradeoff}. Cellular UEs are assumed to transmit with $p_c^k=p_{k,max}^c=200$ mW. For each SE, the corresponding EE is obtained by (\ref{eq:U_EE_D_I}). Simulation results show that the maximum achievable SE and EE decrease monotonically as $I$ increases, which agrees with Corollary 2. In Fig. \ref{EE_SE_tradeoff}, it is impossible to achieve the corresponding EE for some $U_{i, EE}^d$. The reason is that we consider the special case introduced in Section IV that all the signal channels have the same power gain $g$, and all the interference channels have the same power gain $\hat{g}$. In this special case, the channel gains are fixed and no longer depend on the transmission distance. When $I=-15$ dB, $p_c^k=p_{k, max}^c=200$ mW, $N=5, K=3, p_{i, max}^d=200$ mW, the maximum achievable $U_{i, SE}^d$ calculated by (\ref{eq:U_SE_D_I}) is only $8.6182$ bits/s/Hz. Therefore, the solution is infeasible when $U_{i, SE}^d \geq 9$ bits/s/Hz. Both Fig. \ref{EE_SE_real} and Fig. \ref{EE_SE_tradeoff} demonstrate that increasing transmission power beyond the power for optimum EE brings little SE improvement but significant EE loss. However, in the case of $I=-10$dB, the EE loss is not so obvious since that the maximum achievable EE is severely limited by the interference.

\begin{figure}[t]
\begin{center}
\scalebox{0.55} 
{\includegraphics{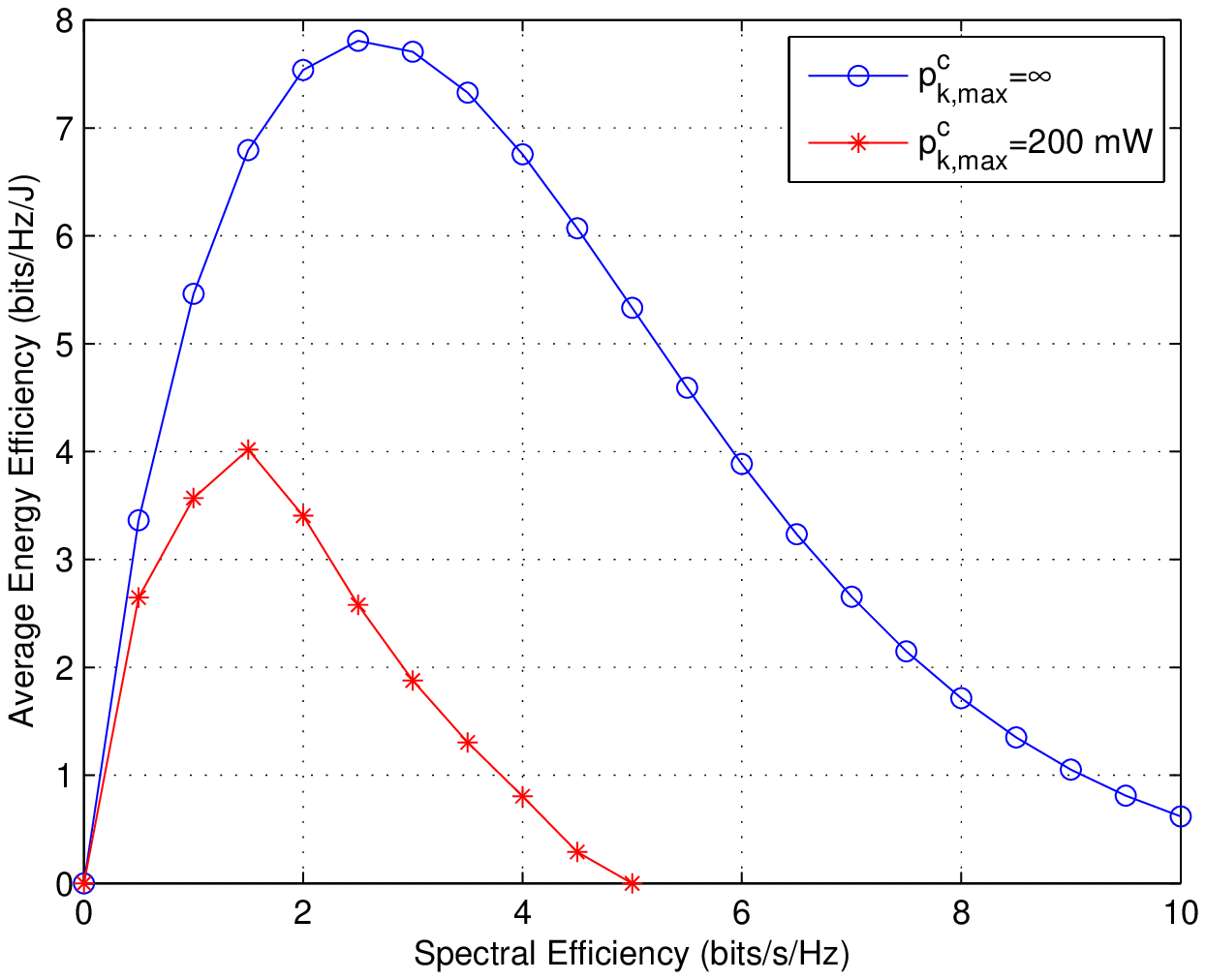}}
\end{center}
\caption{The energy efficiency and spectral efficiency tradeoff for cellular links corresponding to $p_{i,max}^d=\infty$, $200$ mW, ($N=5, K=3$, $N_0=\displaystyle 10^{-7}$ W, $p_{i, max}^d=200$ mW, $\eta=0.35$, $p_{cir}=100$ mW).}
\label{EE_SE_real_cellular}
\end{figure}

\begin{figure}[t]
\begin{center}
\scalebox{0.55} 
{\includegraphics{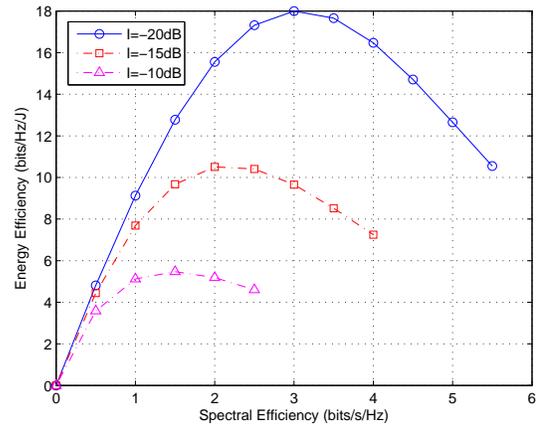}}
\end{center}
\caption{The energy efficiency and spectral efficiency tradeoff for cellular links corresponding to three interference levels $I=-20, -15, -10$ dB, ($g=1, N=5, K=3$, $p_{i, max}^d=p_{k, max}^c=200$ mW, $\eta=0.35$, $p_{cir}=100$ mW).}
\label{EE_SE_tradeoff_cellular}
\end{figure}

Fig. \ref{EE_SE_real_cellular} shows the EE and SE tradeoffs for cellular links corresponding to $p_{i,max}^d=\infty, 200$ mW respectively. The SE requirement is increased from $0$ to $10$ bits/s/Hz with a step of $0.5$,  and the corresponding EE is obtained by Algorithm 1. The average EE of $K$ cellular links is averaged again over a total number of $500$ simulations. Compared with Fig. \ref{EE_SE_real}, the maximum EE is much lower due to the low signal channel gain caused by longer transmission distance in cellular links. In addition, the maximum achievable EE is significantly limited by $p_{k, max}^c$ in low and high SE regimes also due to the long transmission distance. 

 Fig. \ref{EE_SE_tradeoff_cellular} shows the tradeoff between EE and SE for cellular links in the special case discussed in Section \ref{tradeoff}. D2D UEs are assumed to transmit with $p_i^k=\frac{p_{i,max}^d}{K}=\frac{200}{3}$ mW. For each SE, the corresponding EE is obtained by (\ref{eq:U_EE_C_K}). Simulation results show that the maximum achievable SE and EE decrease monotonically as $I$ increases, which agrees with Corollary 3. Compared with Fig. \ref{EE_SE_tradeoff}, both of the maximum EE and SE are limited due to that a cellular link can only use one channel, while a D2D pair uses $K$ channels.

\section{Conclusion}
\label{Conclusion}

In this paper, we proposed a distributed energy-efficient resource allocation algorithm for D2D communications by exploiting the properties of nonlinear fractional programming. We have analyzed and verified the EE and SE tradeoff of the proposed algorithm through computer simulations. Simulation results demonstrate that increasing transmission power beyond the power for optimum EE brings little SE improvement but significant EE loss. Therefore, the proposed energy-efficient algorithm can bring significant EE improvement subject to little SE loss.

\appendices
\section{Proof of the Theorem 1}
\label{theorem1}

The proof of the Theorem 1 is similar to the proof of the Theorem (page 494 in \cite{Dinkelbach}). Firstly, we prove the necessity proof. For any feasible strategy set $\mathbf{p}_i^d$, $\forall i \in \mathcal{N}$, we have 
\begin{align}
\label{eq:sb}
q_i^{d*}=\frac{r_i^d(\mathbf{p}_i^{d*})}{p_{i, total}^d(\mathbf{p}_i^{d*})} \geq \frac{r_i^d(\mathbf{p}_i^d)}{p_{i, total}^d(\mathbf{p}_i^d)}.
\end{align}
By rearranging (\ref{eq:sb}), we obtain
\begin{align}
r_i^d(\mathbf{p}_i^{d*})-q_i^{d*}p_{i, total}^d(\mathbf{p}_i^{d*})&=0,\\
r_i^d(\mathbf{p}_i^{d})-q_i^{d*}p_{i, total}^d(\mathbf{p}_i^{d}) &\leq 0.
\end{align}
Hence, the maximum value of $r_i^d(\mathbf{p}_i^{d})-q_i^{d*}p_{i, total}^d(\mathbf{p}_i^{d})$ is $0$, and can only be achieved by $\mathbf{p}_i^{d*}$, which is obtained by solving the EE maximization problem defined in (\ref{eq:Dproblem}). This completes the necessity proof.

Now we turn to the sufficiency proof. Assume that $\mathbf{\tilde{p}}_i^{d}$ is the optimum solution which satisfies that
\begin{align}
\label{eq:assumption_theorem1}
r_i^d(\mathbf{p}_i^{d})-q_i^{d*}p_{i, total}^d(\mathbf{p}_i^{d}) &\leq r_i^d(\mathbf{\tilde{p}}_i^{d})-q_i^{d*}p_{i, total}^d(\mathbf{\tilde{p}}_i^{d}) =0.
\end{align}
By rearranging (\ref{eq:assumption_theorem1}), we have
\begin{align}
q_i^{d*}=\frac{r_i^d(\mathbf{\tilde{p}}_i^{d})}{p_{i, total}^d(\mathbf{\tilde{p}}_i^{d})} \geq \frac{r_i^d(\mathbf{p}_i^d)}{p_{i, total}^d(\mathbf{p}_i^d)}.
\end{align}
Hence, $\mathbf{\tilde{p}}_i^{d}$ is also the solution of the EE maximization problem defined in (\ref{eq:Dproblem}), i.e., $\mathbf{\tilde{p}}_i^{d}=\mathbf{p}_i^{d*}$. This completes the sufficiency proof.

\section{Proof of the Lemma 1}
\label{lemma1}

It is easily verified that $\frac{\partial U_{i,SE}^d}{\partial p_i^k}=\frac{g_i^k \log_2e}{p_c^k g^k_{c, i}+\sum_{j=1, j\neq i}^{N}p_{j}^k g_{j, i}^k+N_0+p_i^kg_i^k}>0$. Hence, $U_{i,SE}^d$ increases monotonically with $p_i^k$. 

The denominator of $\frac{\partial U_{i,EE}^d}{\partial p_i^k}$ is a positive value, so we only have to consider the numerator, which is defined as 
\begin{align}
f(p_i^k)&=\frac{g_i^k \big (\sum_{k=1}^K \frac{1}{\eta } p_i^k+2p_{cir}\big) \log_2e}{p_c^k g^k_{c, i}+\sum_{j=1, j\neq i}^{N}p_{j}^k g_{j, i}^k+N_0+p_i^kg_i^k}\notag\\
&- \frac{1}{\eta }\sum_{k=1}^K \log_2 \left( 1+\frac{p_i^k g_{i}^k}{p_c^k g^k_{c, i}+\sum_{j=1, j\neq i}^{N}p_{j}^k g_{j, i}^k+N_0} \right)
\end{align}
Take the first-order derivative of $f(p_i^k)$, it can be verified that $\frac{\partial f(p_i^k)}{\partial p_i^k}<0$, thus we have 
$f(\infty) <f(p_i^k)<f(0)$. As $\lim_{p_i^k \to \infty}f(p_i^k)=\frac{1}{\eta } \log_2e - \infty<0$, and $\lim_{p_i^k \to 0}f(p_i^k)=\frac{2 g_i^k p_{cir} \log_2e}{p_c^k g^k_{c, i}+\sum_{j=1, j\neq i}^{N}p_{j}^k g_{j, i}^k+N_0} >0$, we have $\frac{\partial U_{i,EE}^d}{\partial p_i^k}>0$ when $p_i^k<p_i^{k*}$, and $\frac{\partial U_{i,EE}^d}{\partial p_i^k}<0$ when $p_i^k>p_i^{k*}$. Thus, we prove that $U_{i, EE}^d$ increases firstly and then decreases as $p_i^k$ increases. 

Since the numerator and denominator of (\ref{eq:UE_EED}) are concave function and affine function of $p_i^k$ respectively, $U_{i, EE}^d$ is quasiconcave (Problem 4.7 in \cite{convex_optimization}). 

\section{Proof of the Lemma 2}
\label{lemma2}

Taking $U_{i, SE}^d (\mathbf{p}_i^d)-q_i^d p_{i, total}^d(\mathbf{p}_i^d)$ as an example, which is the transformed objective function in subtractive form corresponding to the $i$-th D2D pair. The first part $U_{i, SE}^d (\mathbf{p}_i^d)$ can be rewritten as 
\begin{align}
U_{i, SE}^d (\mathbf{p}_i^d)=\sum_{k=1}^K \log_2 \left( 1+\frac{p_i^k g_{i}^k}{p_c^k g^k_{c, i}+\sum_{j=1, j\neq i}^{N}p_{j}^k g_{j, i}^k+N_0} \right),
\end{align}
which is the sum of $K$ concave functions. The second part $-q_i^d p_{i, total}^d(\mathbf{p}_i^d)$ is given by
\begin{align}
-q_i^d p_{i, total}^d(\mathbf{p}_i^d)=-q_i^d \left( \sum_{k=1}^K \frac{1}{\eta } p_i^k+2p_{cir} \right),
\end{align}
 which is the sum of $K$ affine functions. Since the sum of a concave function and an affine function is also concave, this completes the proof of Lemma 2.
 
 \section{Proof of the Lemma 3}
\label{lemma3}
 
 Define $q_i^{d*}<q_i^{d*'}$, and define $\mathbf{p}_i^{d*}$ and $\mathbf{p}_i^{d*'}$ as the corresponding optimum solutions respectively. We have
\begin{align}
& \max_{(\mathbf{p}_i^d)} U_{i, SE}^d (\mathbf{p}_i^d)-q_i^{d*} p_{i, total}^d(\mathbf{p}_i^d) =  U_{i, SE}^d (\mathbf{p}_i^{d*})-q_i^{d*} p_{i, total}^d(\mathbf{p}_i^{d*}) \notag\\
& >  U_{i, SE}^d (\mathbf{p}_i^{d*'})-q_i^{d*} p_{i, total}^d(\mathbf{p}_i^{d*'})>  U_{i, SE}^d (\mathbf{p}_i^{d*'})-q_i^{d*'} p_{i, total}^d(\mathbf{p}_i^{d*'}) \notag\\
& = \max_{(\mathbf{p}_i^d)} U_{i, SE}^d (\mathbf{p}_i^d)-q_i^{d*'} p_{i, total}^d(\mathbf{p}_i^d). 
\end{align}

 \section{Proof of the Lemma 4}
\label{lemma4}

 Define an feasible solution $\mathbf{\hat{p}}_i^{d}$ such that $q_i^d=\frac{U_{i, SE}^d (\mathbf{\hat{p}}_i^{d})}{p_{i, total}^d (\mathbf{\hat{p}}_i^{d}) }$, we have
\begin{align}
 \max_{\big(\mathbf{p}_i^{d}\big)} U_{i, SE}^d \big(\mathbf{p}_i^{d}\big)-q_i^{d} p_{i, total}^d(\mathbf{p}_i^d )
\geq  U_{i, SE}^d \big(\mathbf{\hat{p}}_i^{d}\big)-q_i^{d} p_{i, total}^d(\mathbf{\hat{p}}_i^d )=0.
\end{align}

 \section{Proof of the Theorem 3}
\label{theorem3}

According to \cite{game_theory_1994}, a Nash equilibrium exists if the utility function is continuous and quasiconcave, and the set of strategies is a nonempty compact convex subset of a Euclidean space. Taking the EE objection function defined in (\ref{eq:UE_EED}) as an example, the numerator $U_{i, SE}^d$ is a concave function of $p_i^k$, $\forall i \in \mathcal{N}, k \in \mathcal{K}$. The denominator $p_{i, total}^d$ is an affine function of $p_i^k$. Therefore, $U_{i, EE}^d$ is quasiconcave (Problem 4.7 in \cite{convex_optimization}). The set of the strategies $\mathbf{p}_i^d=\{p_i^k \mid 0 \leq  \sum_{k=1}^K p_i^k  \leq  p_{i, max}^d, k \in \mathcal{K} \}$, $\forall i \in \mathcal{N}$, is a nonempty compact convex subset of the Euclidean space $\mathbb{R}^K$. Similarly, it is easily proved that the above conditions also hold for the cellular UE. Therefore, a Nash equilibrium exists in the noncooperaive game.

  If the strategy set $\mathbf{p}_i^{d*}$ obtained by using Algorithm \ref{offline algorithm} is not the Nash equilibrium, the $i$-th D2D transmitter can choose the Nash equilibrium $\mathbf{\hat{p}}_i^{d}$ ($\mathbf{\hat{p}}_i^{d} \neq \mathbf{p}_i^{d*}$) to obtain the maximum EE $q_i^{d*}$. However, by Theorem 1, $q_i^{d*}$ can only be achieved by choosing $\mathbf{p}_i^{d*}$. Then, we must have $\mathbf{\hat{p}}_i^{d} = \mathbf{p}_i^{d*}$, which contradicts with the assumption. Therefore, $\mathbf{p}_i^{d*}$ is part of the Nash equilibrium. A similar proof holds for $\mathbf{p}^{c*}_k$. It is proved that the set $\{ \mathbf{p}_i^{d*}, \mathbf{p}^{c*}_k \mid  i \in \mathcal{N},  k \in \mathcal{K}\}$ obtained by using Algorithm \ref{offline algorithm} is the Nash equilibrium.

 \section{Proof of the Theorem 4}
\label{theorem4}

Firstly, we prove that the EE for the $i$-th D2D pair $q_i^d$ increases in each iteration. We denote that $\mathbf{\hat{p}}_i^{d}(n)$ as the optimum resource allocation policies in the $n$-th iteration, and $q_i^{d*}$ as the optimum EE. We denote that $q_i^d (n)$ and $q_i^d (n+1)$ as the EE in the $n$-th iteration and $(n+1)$-th iteration respectively, and we assume that $q_i^d (n) \neq q_i^{d*}$, and $q_i^d (n+1) \neq q_i^{d*}$. $q_i^d (n+1)$ is updated in the $n$-th iteration in the proposed Algorithm 1 as $q_{n+1}=\frac{U_{i, SE}^d \big(\mathbf{\hat{p}}_i^{d}(n)\big)}{p_{i, total}^d \big(\mathbf{\hat{p}}_i^{d}(n)\big) }$. We have
 \begin{align}
 & \max_{\big(\mathbf{p}_i^{d}(n)\big)} U_{i, SE}^d \big(\mathbf{p}_i^{d}(n)\big)-q_i^{d}(n) p_{i, total}^d(\mathbf{p}_i^d (n))\notag\\
&= U_{i, SE}^d \big(\mathbf{\hat{p}}_i^{d}(n)\big)-q_i^d (n) p_{i, total}^d \big( \mathbf{\hat{p}}_i^{d}(n) \big) \notag\\
 &=q_i^d (n+1)p_{i, total}^d \big(\mathbf{\hat{p}}_i^{d}(n)\big)-q_i^d (n) p_{i, total}^d \big( \mathbf{\hat{p}}_i^{d}(n)\big) \notag\\
&=p_{i, total}^d \big(\mathbf{\hat{p}}_i^{d}(n)\big) \big( q_i^d (n+1)-q_i^d (n) \big) \stackrel{\mathrm{Theorem 1, Lemma 3, lemma 4}}{>}0 \notag\\
 &\stackrel{\mathrm{p_{i, total}^d \big(\mathbf{\hat{p}}_i^{d}(n)\big)>0}}{\Longrightarrow }  q_i^d (n+1)>q_i^d (n)   
 \end{align}
 
 Secondly, by combining $q_i^d (n+1)>q_i^d (n)$, Lemma 3, and Lemma 4, we can prove that 
\begin{align}
& \max_{\big(\mathbf{p}_i^{d}\big)} U_{i, SE}^d \big(\mathbf{p}_i^{d}\big)-q_i^{d}(n) p_{i, total}^d(\mathbf{p}_i^d ) \notag\\
&> \max_{\big(\mathbf{p}_i^{d}\big)} U_{i, SE}^d \big(\mathbf{p}_i^{d}\big)-q_i^{d}(n+1) p_{i, total}^d(\mathbf{p}_i^d )\notag\\
& > \max_{\big(\mathbf{p}_i^{d}\big)} U_{i, SE}^d \big(\mathbf{p}_i^{d}\big)-q_i^{d*} p_{i, total}^d(\mathbf{p}_i^d ) \notag\\
&=U_{i, SE}^d \big(\mathbf{p}_i^{d*}\big)-q_i^{d*} p_{i, total}^d(\mathbf{p}_i^{d*} )=0.
\end{align}
Therefore, $q_i^d (n) $ is increased in each iteration and will eventually approaches $q_i^{d*}$ as long as $L_{max}$ is large enough, and $\max_{\big(\mathbf{p}_i^{d}\big)} U_{i, SE}^d \big(\mathbf{p}_i^{d}\big)-q_i^{d} p_{i, total}^d(\mathbf{p}_i^d )$ will approach zero and satisfy the optimality conditions proved in Theorem 1.

 \section{Proof of the Corollary 2}
\label{corollary2}

Since $\frac{\partial U_{i,SE}^d}{\partial I}= - \frac{kp_i^k\log_2e}{\big( p_c^k+(N-1)p_i^k \big)I^2+p_i^kI}<0$, and $\frac{\partial U_{i,EE}^d}{\partial I}= - \frac{k\eta p_i^k\log_2e}{\bigg( \big( p_c^k+(N-1)p_i^k \big)I^2+p_i^kI \bigg)(kp_i^k+2p_{cir}\eta)}<0$, both $U_{i,SE}^d$ and $U_{i,EE}^d$ decreases monotonically as $I$ increases. The second part is proved by setting the numerator of (\ref{eq:U_EE_D_I}) to $0$ and solving the corresponding $U_{i, SE}^d$. 

% or



%\section{Proof of the Theorem 3}
%\label{theorem3}

%Since $U_{i,EE}^d$ is quasiconcave and the set of strategies $\mathbf{p}_i^d$ is a nonempty compact convex subset of the Euclidean space $\mathbb{R}^K$, a Nash equilibrium exists in the noncooperaive game \cite{game_theory_1994}.

  %If the strategy set $\mathbf{p}_i^{d*}$ is not the Nash equilibrium, the Nash equilibrium $\mathbf{\hat{p}}_i^{d}$ ($\mathbf{\hat{p}}_i^{d} \neq \mathbf{p}_i^{d*}$) can be chosen to obtain the maximum EE $q_i^{d*}$. However, by Theorem 1, $q_i^{d*}$ can only be achieved by choosing $\mathbf{p}_i^{d*}$. Then, we must have $\mathbf{\hat{p}}_i^{d} = \mathbf{p}_i^{d*}$, which contradicts with the assumption. Therefore, $\mathbf{p}_i^{d*}$ is part of the Nash equilibrium. A similar proof holds for $\mathbf{p}^{c*}_k$.


\ifCLASSOPTIONcaptionsoff
  \newpage
\fi



%\bibliographystyle{IEEEtran}
\bibliographystyle{IEEEtran}
\bibliography{IEEE_gc_2014}

%\begin{biography}[{\includegraphics[width=1in,height=1.25in,clip,keepaspectratio]{mshell}}]{Michael Shell}


%\newpage


%\vfill


\end{document}